\documentclass[letter, traditabstract]{aa} 
\usepackage{graphicx}
\usepackage[varg]{txfonts}
\usepackage{upgreek} 
\usepackage{multirow} 
\usepackage[FIGTOPCAP, center, nooneline]{subfigure}
\usepackage[]{natbib}

\newcommand{\OI}{O\,{\sc i}}

\bibpunct{(}{)}{;}{a}{}{,}

\begin{document}
   \title{\textit{Trans-cis} molecular photoswitching in interstellar 
   Space\thanks{This paper  makes use of observations obtained with the IRAM-30\,m telescope. IRAM is supported by INSU/CNRS (France), MPG (Germany), and IGN (Spain).}}

     \author{S. Cuadrado\inst{\ref{inst1}}\and J. R. Goicoechea\inst{\ref{inst1}}\and O. Roncero\inst{\ref{inst2}}\and A. Aguado\inst{\ref{inst3}}\and B. Tercero\inst{\ref{inst1}}\and J. Cernicharo\inst{\ref{inst1}}}

   \institute{
   Grupo de Astrof\'{\i}sica Molecular. Instituto de Ciencia de Materiales de Madrid (ICMM-CSIC), Sor Juana Ines de la Cruz 3, E-28049 Cantoblanco, Madrid, Spain.  \email{[s.cuadrado; jr.goicoechea]@icmm.csic.es, jose.cernicharo@csic.es}\label{inst1}    
   \and Instituto de F\'{\i}sica Fundamental (IFF-CSIC). Calle Serrano 123, E-28006 Madrid, Spain. \label{inst2}
   \and Facultad de Ciencias, Unidad Asociada de Qu\'{\i}mica-F\'{\i}sica Aplicada CSIC-UAM, Universidad Aut\'onoma de Madrid, E-28049, Madrid, Spain. \label{inst3}}

   \date{Received October 2016; accepted November 2016}

 
  \abstract{As many organic molecules, formic acid (HCOOH) has two conformers (\textit{trans} and \textit{cis}).
The energy barrier to internal conversion from \textit{trans} to \textit{cis} is much higher than the thermal
energy available in molecular clouds. Thus, only the most stable conformer (\textit{trans}) is expected to exist in detectable amounts. We report the first interstellar detection of \textit{cis}-HCOOH.
Its presence in ultraviolet (UV) irradiated gas exclusively (the Orion Bar photodissociation region), with a low \textit{trans}-to-\textit{cis} abundance ratio of \mbox{2.8 $\pm$ 1.0}, supports a photoswitching mechanism: a given conformer absorbs a stellar photon that radiatively excites the molecule to electronic states above the interconversion barrier. Subsequent fluorescent decay leaves the molecule in a different
conformer form. This mechanism, which we specifically study with \textit{ab initio} quantum calculations, was not considered in Space before but likely induces structural changes of a variety of interstellar molecules submitted to UV radiation.
}

   


   \maketitle
%
\section{Introduction}\label{Intro}

Conformational isomerism refers to isomers (molecules with the same formula but
different chemical structure) having the same chemical bonds but different geometrical
orientations around a single bond. Such isomers are called conformers. An energy barrier often
limits the isomerization. This barrier can be overcome by light. Photoisomerization (or
photoswitching) has been studied in ice IR-irradiation experiments \citep[e.g.][]{Macoas_2004}, in biological processes, and, for
large polyatomic molecules, also in gas-phase experiments \citep{Ryan_2001}. HCOOH is the simplest
organic acid and has two conformers (\textit{trans} and \textit{cis}) depending on the orientation of the
hydrogen single bond. The most stable \textit{trans} conformer was the first acid detected in the
interstellar medium, ISM \citep{Zuckerman_1971}. Gas-phase \textit{trans}-HCOOH shows moderate abundances towards hot cores
\citep{Liu_2001} and hot corinos \citep{Cazaux_2003}, in cold dark clouds \citep{Cernicharo_2012b}, and in cometary coma \citep{Bockelee_2000}. 
Solid HCOOH is present in interstellar ices \citep{Keane_2001} and in chondritic meteorites \citep{Briscoe_1993}.

The ground-vibrational state of \textit{cis}-HCOOH is \mbox{1365 $\pm$ 30 cm$^{-1}$} higher in energy than the
\textit{trans} conformer \citep{Hocking_1976}. The energy barrier to internal rotation (the conversion from \textit{trans} to \textit{cis}) is about 4827\,cm$^{-1}$ \citep{Hocking_1976}, approximately 7000\,K in temperature units. This is much higher than the thermal energy available in molecular clouds (having typical temperatures from about 10 to 300\,K). Generalizing this reasoning, only the most stable conformer of a given species would be
expected in such clouds. Photoswitching, however, may be a viable mechanism producing the
less stable conformers in detectable amounts: a given conformer absorbs a high-energy photon
that radiatively excites the molecule to electronic states above the interconversion energy barrier.
Subsequent radiative decay to the ground-state would leave the molecule in a different
conformer.

In this work we have searched for pure rotational lines of the \textit{trans}- and \textit{cis}-HCOOH
conformers in the 3\,millimetre spectral band. We observed three prototypical interstellar sources
known to display a very rich chemistry and bright molecular line emission: (i) the Orion Bar photodissociation region (PDR): the edge of the Orion cloud irradiated by ultraviolet (UV) photons from nearby massive stars \citep[e.g.][]{Goicoechea_2016},
(ii) the Orion hot core: warm gas around massive protostars \citep[e.g.][]{Tercero_2010}, and (iii) Barnard 1-b (B1-b): a
cold dark cloud \citep[e.g.][]{Cernicharo_2012b}. The two latter sources are shielded from strong UV radiation
fields. We only detect \textit{cis}-HCOOH towards the Orion Bar. This represents the first interstellar detection of the conformer.

\begin{figure*}[th]
\centering
\includegraphics[scale=0.52,angle=0]{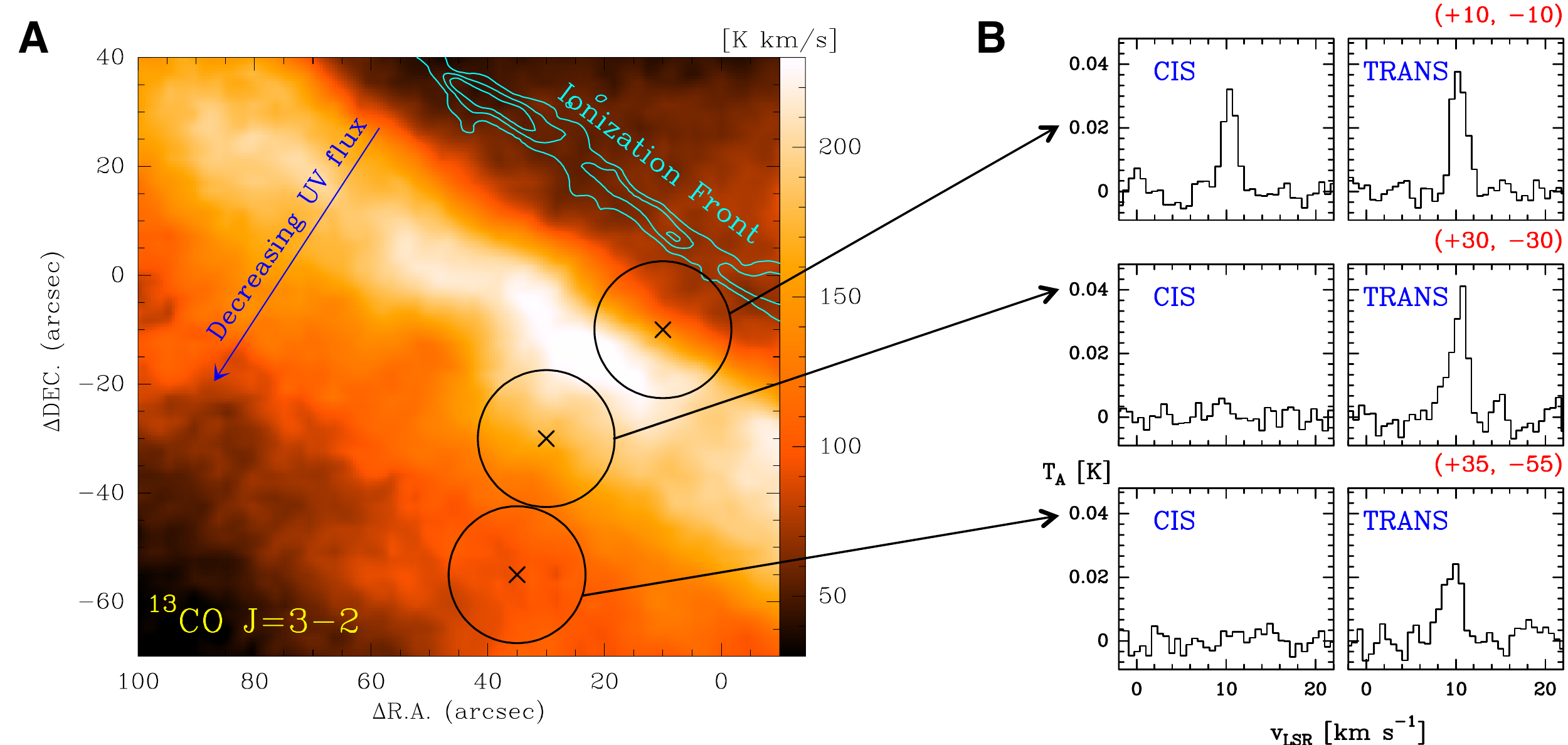}
\caption{Detection of \textit{cis}-HCOOH towards the FUV-illuminated edge of the Orion Bar.
\textit{Left}: $^{13}$CO \mbox{$J$ = 3 $\rightarrow$ 2} integrated emission image with
a HPBW of 8$''$ obtained
with the IRAM-30\,m telescope (Cuadrado et al. in prep.). The cyan contour marks the position of neutral cloud boundary traced by the \OI\, 1.317 $\upmu$m fluorescent line emission \citep[in contours from 3 to 7 by \mbox{2 $\times$ 10$^{-4}$ erg s$^{-1}$ cm$^{-2}$ sr$^{-1}$;}][]{Walmsley_2000}. \textit{Right}: \textit{Cis}- and \textit{trans}-HCOOH stacked spectra towards the observed positions.}
\label{fig:map_stacking}
\end{figure*}

\section{Source selection and observations}\label{Obs}

Because of its nearly edge-on orientation, the Orion Bar PDR is a template source to study the
molecular content as the far-UV radiation field (FUV; stellar photons with energies below
13.6\,eV, or wavelengths ($\uplambda$) longer than 911\,$\AA$, the hydrogen atom ionisation threshold) is
attenuated from the cloud edge to the interior \citep{Hollenbach_1999}. The impinging FUV radiation field at the
edge of the Bar is about \mbox{4 $\times$ 10$^{4}$} times the mean interstellar radiation field \citep[e.g.][and references therein]{Goicoechea_2016}. We observed three positions of the Bar characterized by a decreasing FUV photon flux.

We have used the IRAM-30\,m telescope (Pico Veleta, Spain) and the 90\,GHz EMIR receiver. We employed the Fast Fourier Transform Spectrometer
(FFTS) backend at 200\,kHz spectral resolution \mbox{(0.7 km s$^{-1}$} at 90\,GHz). 
Observations towards the Orion Bar are part of a complete millimetre (mm) line survey \mbox{(80 $-$ 360 GHz,} \citealt{Cuadrado_2015a}). They include specific deep searches for HCOOH lines in the 3\,mm band towards three different
positions located at a distance of 14$''$, 40$''$, and 65$''$ from the ionisation front
(Fig.~\ref{fig:map_stacking}A). 
Their offset coordinates with respect to the 
\mbox{$\mathrm{\alpha_{2000}=05^{h}\,35^{m}\,20.1^{s}\,}$}, 
\mbox{$\mathrm{\delta_{2000}=-\,05^{\circ}25'07.0''}$}
position at the ionisation front are \mbox{(+10$''$, -10$''$)}, \mbox{(+30$''$, -30$''$')}, and \mbox{(+35$''$, -55$''$)}. 
The observing procedure was position switching with a reference position at \mbox{($-$600$''$, 0$''$)} to avoid the extended
emission from the Orion molecular cloud. 
The half power beam width (HPBW) at 3\,mm ranges from \mbox{$\sim$30.8$''$ to 21.0$''$}.
We reduced and analysed the data using the GILDAS software 
as described in \citet{Cuadrado_2015a}. 
The antenna temperature, $T^{*}_{\rm A}$, was
 converted to the main beam temperature, $T_{\rm MB}$, using \mbox{$T_{\rm MB}$ = $T^{*}_{\rm A}/ \upeta_{\rm MB}$}, where $\upeta_{\rm MB}$ is the antenna efficiency \mbox{($\upeta_{\rm MB}$ = 0.87 $-$ 0.82} at 3\,mm). 
The rms noise obtained after 5\,h integration is  \mbox{$\sim$1 $-$ 5\,mK} per resolution channel.

We also searched for HCOOH in regions shielded from strong
FUV radiation fields (see Appendix~\ref{non-detection}). We selected two chemically rich sources for which
 we have also carried out deep mm-line surveys with
the IRAM-30m telescope: towards the hot core in Orion BN/KL (Tercero et al.
2010) and towards the quiescent dark cloud Barnard 1-b (B1-b;
Cernicharo et al. 2012).

\section{Results}

\subsection{Line identification}\label{indetification}

We specifically computed the \textit{cis}-HCOOH rotational lines frequencies by fitting the available laboratory data \citep{Winnewisser_2002} with our own spectroscopic code, MADEX \citep{Cernicharo_2012}. The standard
deviation of the fit is 60\,kHz. For the \textit{trans} conformer, higher frequency laboratory data \citep{Cazzoli_2010}
were also used in a separate fit. The standard deviation of the fit for \textit{trans}-HCOOH is 42\,kHz.
These deviations are smaller than the frequency resolution of the spectrometer we used to carry
out the astronomical observations. Formic acid is a near prolate symmetric molecule with
rotational levels distributed in different $K_{\rm a}$ rotational ladders \mbox{($K_{\rm a}$ = 0,} 1, 2...). 
Both $a$- and $b$-components of its electric dipole moment $\upmu$ exist \citep{Winnewisser_2002}. 
The dipole moments of the \textit{cis} conformer
\mbox{($\upmu_{\rm a}$ = 2.650 D} and \mbox{$\upmu_{\rm b}$ = 2.710 D,} \citealt{Hocking_1976}) are stronger than those of the \textit{trans} conformer
\mbox{($\upmu_{\rm a}$ = 1.421 D} and \mbox{$\upmu_{\rm b}$ = 0.210 D,} \citealt{Kuze_1982}). 

\begin{figure}
\centering
\includegraphics[scale=0.4,angle=0]{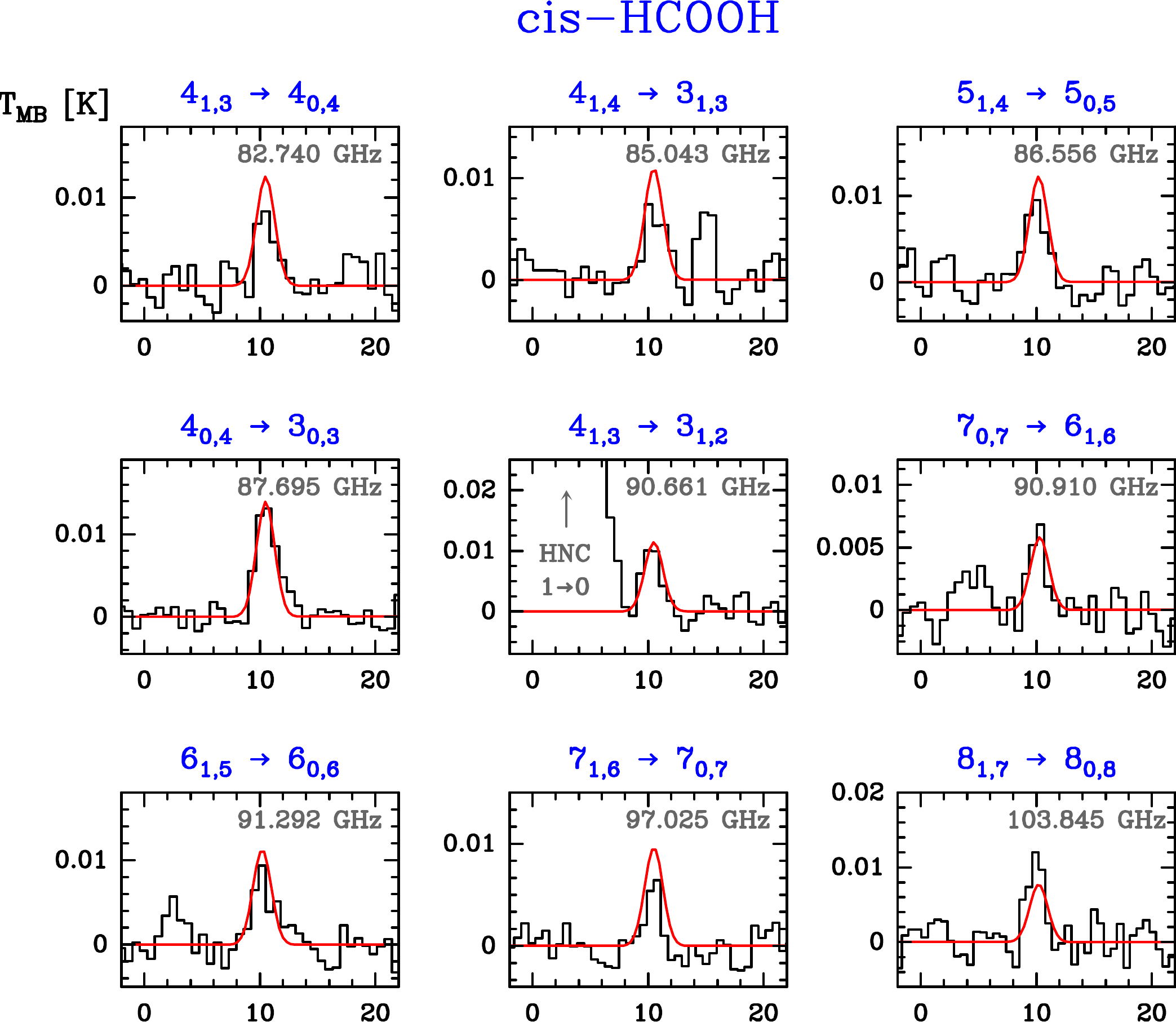}\\
\vspace{0.45cm}
\includegraphics[scale=0.4,angle=0]{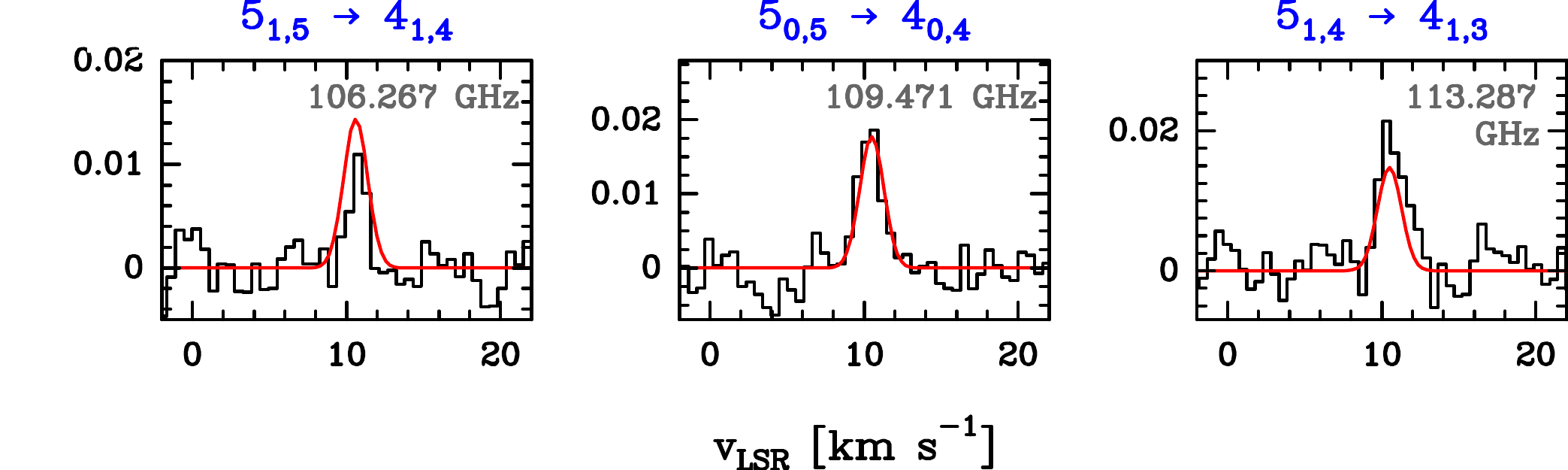}
\caption{Detected \textit{cis}-HCOOH rotational lines towards the Orion Bar, \mbox{(+10$''$, -10$''$)} position. The ordinate refers to
the intensity scale in main beam temperature units, and the abscissa to the LSR
velocity. Line frequencies (in GHz) are indicated at the top-right of each panel together with the
rotational quantum numbers (in blue). The red curve shows an excitation model that reproduces
the observations.}
\label{fig:cis_OB}
\end{figure}

In total, we identify 12 rotational lines of \textit{cis}-HCOOH and 10 of \textit{trans}-HCOOH above 3$\sigma$ towards the FUV-illuminated edge of the Orion Bar, \mbox{(+10$''$, -10$''$)} position. 
The detected lines from the \textit{cis}- and \textit{trans}-HCOOH are
shown in Figs.~\ref{fig:cis_OB} and \ref{fig:trans_OB}, respectively. Lines attributed to HCOOH show a Gaussian line profile centred at the systemic velocity of the Orion Bar \mbox{(10.4 $\pm$ 0.3 km s$^{-1}$)}. Lines are narrow, with linewidths of \mbox{1.9 $\pm$ 0.3 km s$^{-1}$}. 
The large number of detected lines, and the fact that none of the
lines correspond to transitions of abundant molecules known to be present in the Bar or in
spectroscopic line catalogues, represents a robust detection of the \textit{cis} conformer.
 The observational parameters and Gaussian
fit results are tabulated in Tables~\ref{Table_cisOB} and \ref{Table_transOB} for the \textit{cis} and \textit{trans} conformer, respectively.

\subsection{Line stacking analysis}

Complex organic molecules have relatively low abundances in FUV-irradiated interstellar
gas \citep{Guzman_2014}. Indeed, detected \textit{trans}-HCOOH lines are faint. To improve the statistical significance
of our search towards the positions inside the Bar, we performed a $``$line stacking$"$ analysis. For 
each observed position, we added spectra at the expected frequency of several
HCOOH lines that could be present within the noise level (sharing similar rotational level
energies and Einstein coefficients). The spectra in frequency scale were first converted to local standard of rest (LSR) velocity scale and resampled to the same velocity channel resolution before stacking.
We repeated this procedure for \textit{trans}-HCOOH lines. This
method allows us to search for any weak line signal from the two conformers that could not be
detected individually. 

Figure~\ref{fig:map_stacking}B shows a comparison of the stacking results for \textit{cis} and \textit{trans}-HCOOH lines towards the three target positions in the Bar. Although we detect \textit{trans}-HCOOH in all positions, emission from the \textit{cis} conformer is only detected towards the position located closer to the cloud edge, \mbox{(+10$''$, -10$''$)}. They demonstrate that \textit{cis}-HCOOH is detected close the FUV-illuminated edge of the Bar, but the emission disappears towards the more shielded cloud interior.

A similar stacking analysis was carried out for the Orion
hot core and B1-b spectra. Although we detect several \textit{trans}-HCOOH lines, the \textit{cis} conformer
was not detected towards the hot core and the cold dark cloud (see Appendix~\ref{non-detection}).

\subsection{\textit{Trans}-to-\textit{cis} abundance ratios}

Given the number of HCOOH lines detected towards the Bar, we can determine the
column density and rotational temperatures of both conformers accurately (see Appendix~\ref{DR}). In particular, we infer a low \textit{trans}-to-\textit{cis}
abundance ratio of \mbox{2.8 $\pm$ 1.0}. The non-detection of \textit{cis}-HCOOH towards the Orion hot core and
\mbox{B1-b} (see Appendix~\ref{non-detection}) provides much higher \textit{trans}-to-\textit{cis} limits ($>$100 and $>$60, respectively). This suggests that the presence of \textit{cis}-HCOOH in the Orion Bar PDR is related to the strong FUV field
permeating the region.

\section{Photoisomerization rates and discussion} \label{Discussion}

Photolysis of HCOOH has been widely studied, both experimentally \citep{Sugarman_1943,Ioannoni_1990,Brouard_1992,Su_2000} and
theoretically \citep{Beaty-Travis_2002,He_2003,Maeda_2015}. 
Dissociation of HCOOH takes place after absorption of FUV photons
with energies greater than \mbox{$\sim$5\,eV} \mbox{($\uplambda$ $<$ 2500\,$\AA$)}. Recently, \citet{Maeda_2015} determined that this dissociation threshold coincides with the crossing of the S$_0$ and T$_1$
electronic states of the molecule. The specific products of the photofragmentation process (of the
different photodissociation channels) depend on the specific energy of the FUV photons and on
the initial HCOOH conformer. Interestingly, absorption of lower energy photons does not
dissociate the molecule but induces fluorescent emission. In particular, HCOOH fluorescence
from the S$_1$ excited electronic state has been observed in laser-induced experiments performed in
the \mbox{$\uplambda$ = 2500 $-$ 2700 $\AA$} range \citep{Ioannoni_1990,Brouard_1992}. These studies indicate that the 
geometrical configuration of
the two hydrogen atoms is different in the S$_0$ and S$_1$ states. The fluorescence mechanism from the
S$_1$ state is a likely route for the \mbox{\textit{trans} $\rightarrow$ \textit{cis}} isomerization. In addition, the isomerization barrier from the S$_1$ state \mbox{($\sim$1400\,cm$^{-1}$)} is much lower than from the ground.

\begin{figure}[t]
\centering
\includegraphics[scale=0.5,angle=0]{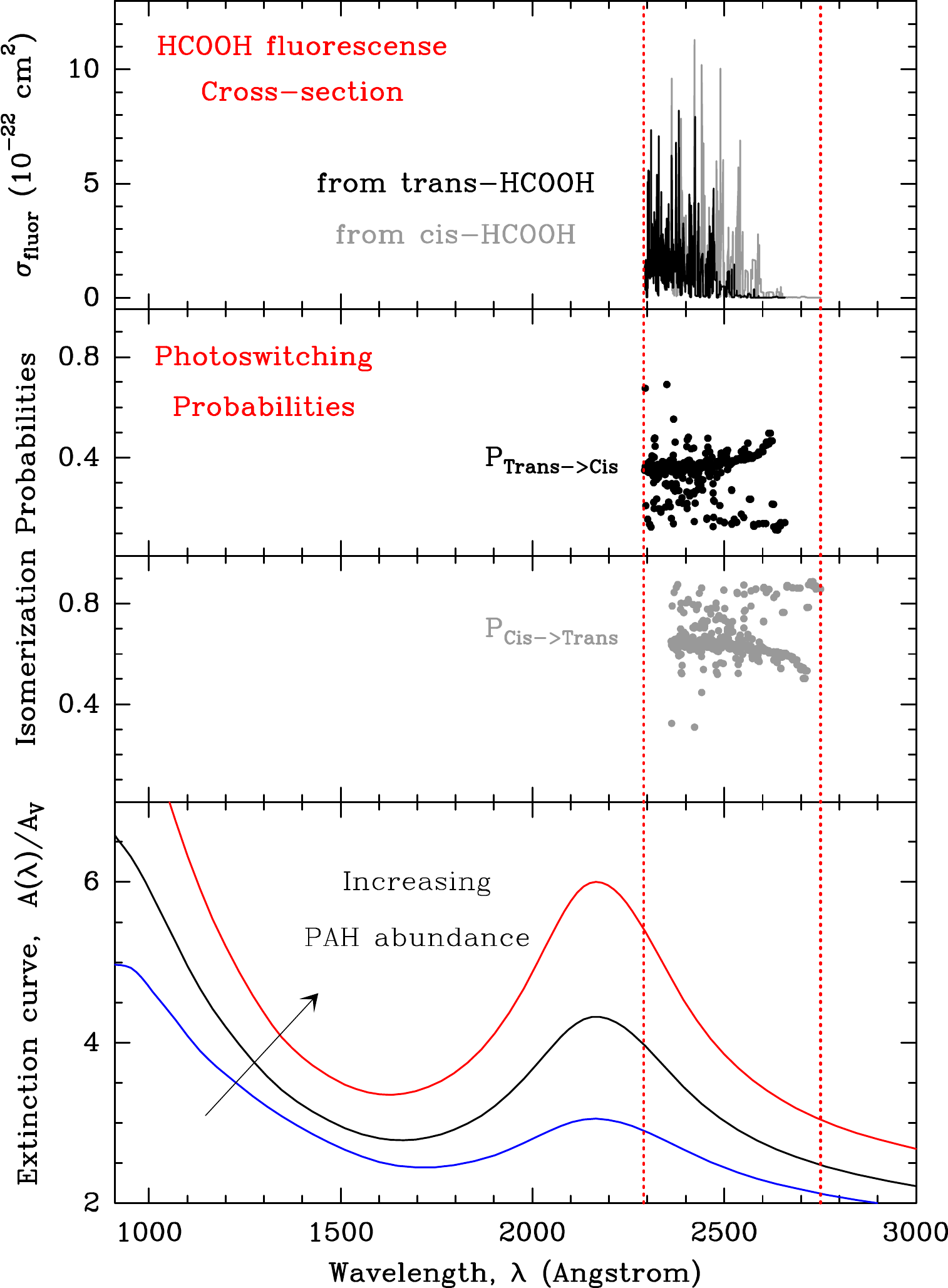}
\caption{\textit{Ab initio} absorption cross-sections and photoisomerization probabilities computed in this work. 
\textit{Top panel}: \textit{trans}- and \textit{cis}-HCOOH absorption cross-sections for photons with \mbox{$E$ $<$ 5 eV} (those leading to fluorescence).
 \textit{Middle panels}: normalized probabilities of bound-bound decays producing
isomerization (\textit{trans $\rightarrow$ cis} and \textit{cis $\rightarrow$ trans}). 
\textit{Bottom panel}: standard interstellar dust extinction
curve  (blue). Black and red curves show the effect of an increased PAH abundance.}
\label{fig:cross_extin}
\end{figure}

In order to quantify the role of the photoswitching mechanism, we carried out \textit{ab initio}
quantum calculations and determined the HCOOH potential energy surfaces of the S$_0$ and S$_1$
electronic states as a function of the two most relevant degrees of freedom, $\upphi_1$ the torsional angle
of OH and $\upphi_2$, the torsional angle of CH (see Appendix~\ref{Ab_initio_estimation} and Fig.~\ref{fig:pot_energ}). With this
calculation we can compute the position of the photon absorptions leading to fluorescence (those
in the approximate \mbox{$\uplambda$ = 2300 $-$ 2800 $\AA$} range), and the probabilities to fluoresce from one
conformer to the other (the \textit{trans}-to-\textit{cis} and \textit{cis}-to-\textit{trans} photoswitching cross-sections and probabilities, see Fig.~\ref{fig:cross_extin}).

With a knowledge of $N_{\rm ph}(\uplambda)$, the FUV photon flux in units of photon
 \mbox{cm$^{-2}$ s$^{-1}$ $\AA^{-1}$}, we can calculate the number of \textit{trans}-to-\textit{cis} and \textit{cis}-to-\textit{trans} photoisomerizations per second ($\upxi_{\rm tc}$ and $\upxi_{\rm ct}$,
respectively. See Appendix~\ref{photoisomerization_rate}). In the absence of any other mechanism destroying HCOOH, the \mbox{$\upxi_{\rm ct}$/$\upxi_{\rm tc}$} ratio provides the \textit{trans}-to-\textit{cis} abundance ratio in equilibrium. The time needed to reach the
equilibrium ratio is then \mbox{($\upxi_{\rm tc}$ + $\upxi_{\rm ct}$)$^{-1}$}. $N_{\rm ph}(\uplambda)$, and thus $\upxi_{\rm tc}$ and $\upxi_{\rm ct}$, depend on the FUV radiation sources (type of star) and on the cloud position. Describing the cloud depth position in terms of the visual extinction into the cloud ($A_{\rm V}$), one magnitude of extinction is equivalent to a column density of about 10$^{21}$ H$_2$ molecules per cm$^{-2}$ in the line-of-sight.

In general (for a flat, wavelength-independent FUV radiation field), HCOOH
photodissociation will always dominate over fluorescence (photodissociation cross-sections are
larger and the relevant photons can be absorbed over a broader energy range, \mbox{$E$ $>$ 5\,eV}). The
strength and shape of the interstellar FUV radiation field, however, is a strong function of $A_{\rm V}$ and
is very sensitive to the dust and gas absorption properties. Because of the wavelength-dependence of the FUV-absorption process, $N_{\rm ph}(\uplambda)$ drastically changes as one moves from the
cloud edge to the shielded interior. In particular, the number of low-energy FUV photons
(e.g. below 5\,eV) relative to the high-energy photons (e.g. those above 11\,eV dissociating
molecules such as CO and ionising atoms such as carbon) increases with $A_{\rm V}$. 
In this work we used a
FUV radiative transfer and thermo-chemical model \citep{LePetit_2006,Goicoechea_2007} to estimate $N_{\rm ph}(\uplambda)$ at different positions
of the Orion Bar. The well-known $``$2175 $\AA$ bump$"$ of the dust extinction curve (absorption of
\mbox{$\uplambda$ = 1700 $-$ 2500\,$\AA$} photons by PAHs and small carbonaceous grains, \citealt{Cardelli_1989,Joblin_1992}) greatly reduces the number of
HCOOH dissociating photons relative to those producing HCOOH fluorescence (Fig.~\ref{fig:cross_extin}, bottom panel). The
resulting FUV radiation spectrum, $N_{\rm ph}(\uplambda)$, at different $A_{\rm V}$ is used to calculate $\upxi_{\rm tc}$ and $\upxi_{\rm ct}$ (Table~\ref{Table_photo_rates}). We determine that at a cloud depth  of about \mbox{$A_{\rm V}$ = 2 $-$ 3 mag}, and if the
number of HCOOH dissociating photons is small compared to the number of photons producing
photoisomerization (i.e. most photons with \mbox{$E$ $>$ 5\,eV} have been absorbed), the \textit{cis} conformer
should be detectable with a \textit{trans}-to-\textit{cis} abundance ratio of about \mbox{3.5 $-$ 4.1}. These values are remarkably close to the \textit{trans}-to-\textit{cis} ratio inferred from our observations of the Bar.

Closer to the irradiated cloud edge (\mbox{$A_{\rm V}$ = 0 $-$ 2 mag}), photodissociation destroys the
molecule much faster than the time needed for the \textit{trans}-to-\textit{cis} isomerization. On the other hand,
too deep inside the cloud, the flux of \mbox{$E$ $>$ 5\,eV} photons decreases to values for which the
isomerization equilibrium would take an unrealistic amount of time ($>$10$^{6}$ years for \mbox{$A_{\rm V}$ = 5 mag}). Therefore, our detection of \textit{cis}-HCOOH in irradiated cloud layers where CO
becomes the dominant carbon carrier (a signature of decreasing flux of high-energy FUV
photons) agrees with the photoswitching scenario.

For standard  grain properties and neglecting HCOOH photodissociation, we
calculate that the time needed to achieve a low \textit{trans}-to-\textit{cis} abundance ratio and make
\textit{cis}-HCOOH detectable at \mbox{$A_{\rm V}$ = 2 $-$ 3 mag} is \mbox{10$^{4}$ $-$ 10$^{5}$} years (see Table~\ref{Table_photo_rates}). 
This is reasonably fast, and
shorter than the cloud lifetime. In practice, it is not straightforward to quantify the exact
contribution of HCOOH photodissociation and photoisomerization at different cloud positions.
The above time-scales require that the flux of \mbox{$E$ $>$ 5\,eV} dissociating photons is small compared to
those producing fluorescence. This depends on the specific dust absorption properties, that
sharply change with $A_{\rm V}$ as dust populations evolve \citep{Draine_2003}, on the strength and width of the 2175\,$\AA$
extinction bump, and on the role of molecular electronic transitions blanketing the FUV
spectrum. The similar \textit{trans}-HCOOH line intensities observed towards the three positions of the
Bar (Fig.~\ref{fig:map_stacking}) suggest that even if the HCOOH photodestruction rate increases at the irradiated
cloud edge, the HCOOH formation rate (from gas-phase reactions or desorbing directly from
grain surfaces, \citealt{Garrod_2008}) must increase as well. The inferred HCOOH abundances are not particularly
high, \mbox{(0.6 $-$ 3.0) $\times$ 10$^{-10}$} with respect to H. Hence, modest HCOOH photodestruction and
formation rates are compatible with the photoswitching mechanism occurring in realistic
times. 

Although the observed abundances of \textit{trans}- and \textit{cis}-HCOOH in the Orion Bar are compatible with gas-phase photoisomerization, we note that photoswitching may also occur on the surface of grains covered by HCOOH ices. 
In a similar way, solid HCOOH  (mostly \textit{trans}) can absorb FUV photons that switch the molecule to the \textit{cis} form before being desorbed. Once in the gas, both conformers will continue their photoisomerization following absorption of \mbox{$\lambda$ $\gtrsim$ 2500\,$\AA$} photons. Laboratory experiments are needed to quantify the mechanisms leading to HCOOH ice photoswitching by FUV photon absorption.  

Searching for further support to the FUV photoswitching scenario, we qualitatively explored two
other possibilities for the \textit{trans}-to-\textit{cis} conversion. First, the
isomerization of solid HCOOH after IR irradiation of icy grain surfaces (as observed in the
laboratory, \citealt{Macoas_2004,Olbert-Majkut_2008}) and subsequent desorption to the gas-phase. Second, the gas-phase isomerization
by collisions of HCOOH with energetic electrons \mbox{($\sim$0.5\,eV)}. 
We concluded that if these were the
dominant isomerization mechanisms, emission lines from \textit{cis}-HCOOH would have been detected in other interstellar sources (see Appendix~\ref{alternative_mechan}).

Isomerization by absorption of UV photons  was not considered as a possible
mechanism to induce structural changes of molecules in interstellar gas. The detection of
\textit{cis}-HCOOH towards the Orion Bar opens new avenues to detect a variety of less stable
conformers in Space. This can have broad implications in astrochemistry and astrobiology.


\begin{acknowledgements} 
We thank N. Marcelino for helping with the observations of B1-b. We
thank the ERC for support under grant ERC-2013-Syg-610256- NANOCOSMOS. We also thank
Spanish MINECO for funding support under grants AYA2012-32032 and FIS2014-52172-C2,
and from the CONSOLIDER-Ingenio program $``$ASTROMOL$"$ CSD 2009-00038. IRAM is
supported by INSU/CNRS (France), MPG (Germany), and IGN (Spain).
\end{acknowledgements}


\bibliographystyle{aa}
\bibliography{references}


\appendix

\section{\textit{Ab initio} estimation of fluorescence cross-sections and photoisomerization probabilities} \label{Ab_initio_estimation}

\begin{figure}
\centering
\includegraphics[scale=0.8,angle=0]{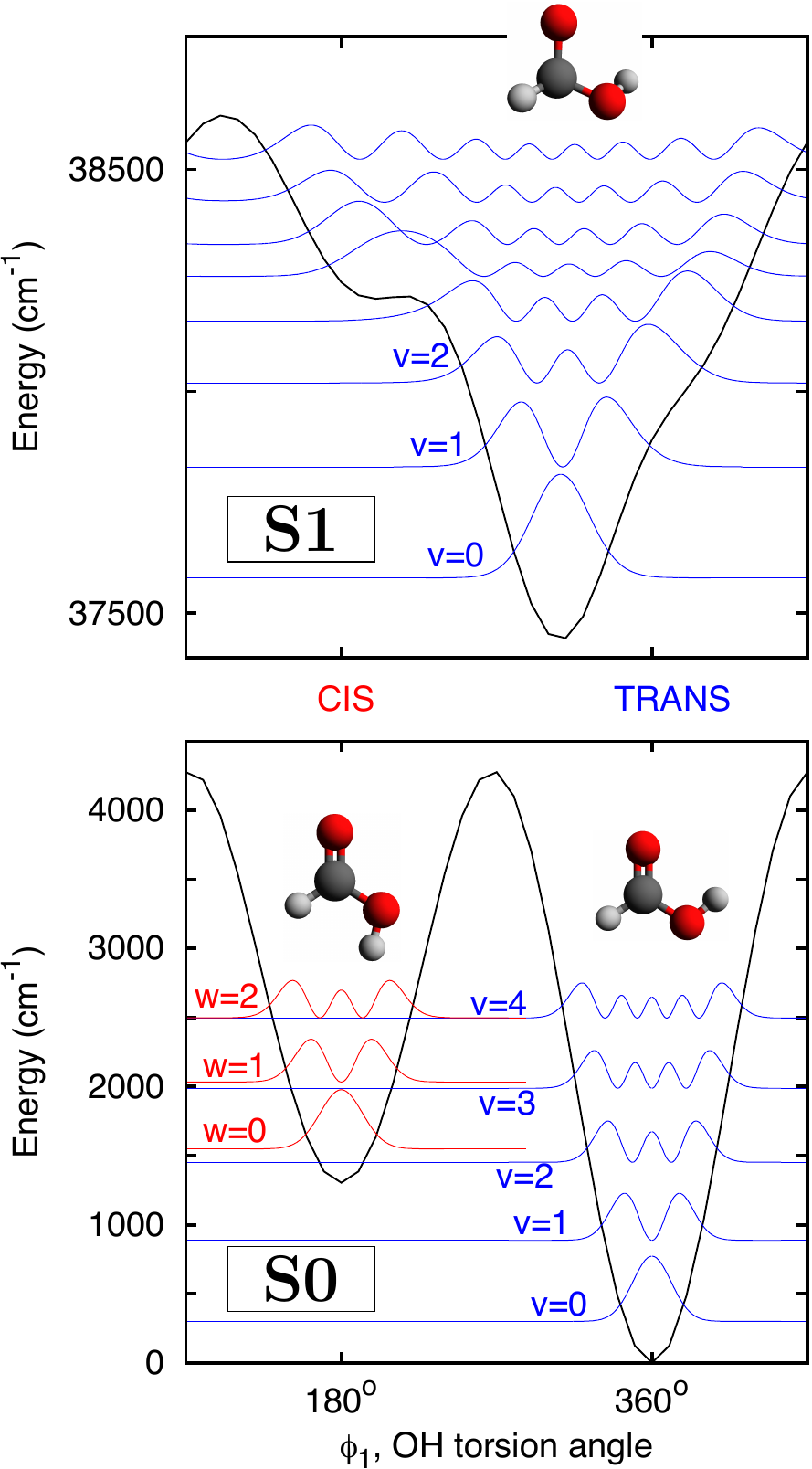}
\caption{One-dimensional potential energy surfaces of HCOOH as function of the OH
torsional angle $\upphi_1$. \textit{Bottom panel}: ground S$_{0}$ electronic state. \textit{Top panel}: excited S$_{1}$ state. One-dimensional cuts were obtained from the two-dimensional grid (see text) by setting \mbox{$\upphi_2$ = 180$^{\circ}$} and \mbox{$\upphi_2$ = 300$^{\circ}$} for S$_{0}$ and S$_{1}$, respectively. 
We also show the vibrational-wave functions obtained from a one-dimensional model. The different geometrical structures of the HCOOH molecule in each energy minimum are shown.}
\label{fig:pot_energ}
\end{figure}

In this appendix, we demonstrate that the detected \textit{cis}-HCOOH towards the Orion Bar can
be produced by a gas-phase photoswitching mechanism. To estimate the FUV photon absorption cross-sections
 and probabilities of the \textit{trans-cis} photoisomerization process, we start calculating the
potential energy surfaces of the HCOOH S$_{0}$ and S$_{1}$ electronic states as a function of the two most
relevant degrees of freedom, the torsional angle of OH ($\upphi_1$), and the torsional angle of CH ($\upphi_2$)
(see Fig.~\ref{fig:pot_energ}).

We performed ic-MRCI-F12 \textit{ab initio} calculations using the MOLPRO suite of programs\footnote{MOLPRO \citep{MOLPRO}, version 2012, is a package of \textit{ab initio} programs for advanced molecular electronic structure calculations, designed and maintained by H.-J. Werner and P. J. Knowles, and with contributions from many other authors (see http://www.molpro.net).} with the VDZ-F12 basis set. The obtained results agree with the stationary points previously
reported by \citet{Maeda_2012,Maeda_2015}. The molecular orbitals and reference configurations were determined with a
CASSCF calculation using 16 active orbitals. The optimized equilibrium geometries in the S$_{0}$
and S$_{1}$ electronic states are in agreement with previous results, corresponding to planar and bent
\textit{trans}-HCOOH conformers, respectively. They are listed in Table~\ref{Table_geom_trans}. For \textit{trans}-HCOOH, the
normal modes in the S$_{0}$ state have the following frequencies: 628.59, 662.86, 1040.64, 1117.90,
1316.0, 1416.22, 1792.32, 3083.01, and 3749.88 cm$^{-1}$. The two lowest frequencies correspond to
the torsional angles of the OH and CH bonds, respectively.

For the two lower singlet states, S$_{0}$ and S$_{1}$, we calculate a two-dimensional grid composed of
37 equally spaced points for $\upphi_1$ and $\upphi_2$, fixing the rest of coordinates to the corresponding values
listed in Table~\ref{Table_geom_trans}. These points are interpolated using a two-dimension splines method to get the
potential energy surfaces, S$_{0}$ and S$_{1}$, at any desired geometry, including the two conformers.

The potential energy surface of the S$_{0}$ electronic state presents two minima for
 \mbox{$\upphi_2$ = 0$^{\circ}$}, one at \mbox{$\upphi_1$ = 0$^{\circ}$} or 360$^{\circ}$ (\textit{trans}), 
 and a second at \mbox{$\upphi_1$ = 180$^{\circ}$} (\textit{cis}). As illustrated in Fig.~\ref{fig:pot_energ}, both minima
correspond to a planar geometry. The potential for the S$_{1}$ excited state presents two equivalent
wells for the \textit{trans}-conformer \mbox{($\upphi_1$ = 300$^{\circ}$}, \mbox{$\upphi_2$ = 120$^{\circ}$} or
\mbox{$\upphi_1$ = 60$^{\circ}$}, \mbox{$\upphi_2$ = 240$^{\circ}$)}. Therefore, the
minimum geometrical configuration in the S$_{1}$ excited state is no longer planar. The \textit{cis} conformer
minimum transforms into a shoulder of the potential. This is shown in the one-dimensional cut
shown in Fig.~\ref{fig:pot_energ} for the case of \mbox{$\upphi_2$ = 120$^{\circ}$}. In this case, the potential energy surface as a function
of $\upphi_1$ is rather flat, while it shows a double-well structure as a function of $\upphi_2$, corresponding to
geometries above and below the molecular plane.

 \begin{table}[ht]
 \centering 
 \caption{Optimized geometries for \textit{trans}-HCOOH in the ground (S$_{0}$) and excited
electronic state (S$_{1}$). Distances are in Angstrom and angles in degree units.}
 \label{Table_geom_trans}     
  	
 \begin{tabular}{l c c@{\vrule height 10pt depth 5pt width 0pt}}     
 \hline\hline      

 Geometry of            & S$_{0}$ ground-state  & S$_{1}$ excited-state  \\
 \textit{trans}-HCOOH            & (geom-S$_{0}$)        &    (geom-S$_{1}$)      \\

  \hline                                                                    
\ \ R(CO$_{1}$)                             &  1.1987            &           1.3683     \\
\ \ R(CO$_{2}$)                             &  1.3600            &           1.3840     \\
\ \ $\uptheta$(O$_{1}$CO$_{2}$)             &  122.38            &           111.17     \\
\ \ R(CH$_{1}$)                             &  1.1008            &           1.0751     \\
\ \ $\uptheta$(O$_{1}$CH$_{1}$)             &  124.01            &           113.06     \\
\ \ R(OH$_{1}$)                             &  0.9663            &           0.9661     \\
\ \ $\uptheta$(CO$_{2}$H$_{2}$)             &  108.72            &           107.51     \\
\ \ $\upphi_1$(H$_{2}$O$_{2}$CO$_{1}$)      &  0.00              &           55.32      \\
\ \ $\upphi_2$(H$_{1}$CO$_{1}$O$_{2}$)      &  180.00            &           232.28     \\
      \hline                                                                        
               
        \end{tabular}   
        \end{table}

We solved the two-dimension Shr\"odinger equation for $\upphi_1$ and $\upphi_2$ and obtained the vibrational
eigenfunctions. The first six vibrational levels of the S$_{0}$ electronic state correspond to the \textit{trans}
conformer, the seventh energy level corresponds to the ground-vibrational state of \textit{cis}-HCOOH.
In the S$_{1}$ excited electronic state, the presence of a double well as a function of $\upphi_2$ implies that
two degenerate vibrational states appear. The two well depths are different in geom-S$_{0}$ and
geom-S$_{1}$ which means that their nodal structure significantly changes.

In a second step, we calculate the transition dipole moments for the two-dimensional grids of
geom-S$_{0}$ and geom-S$_{1}$, and determine the transitions between the S$_{0}$ state and the S$_{1}$ state. We
derive the absorption spectra starting from both \textit{trans}-HCOOH \mbox{($\upnu$ = 0)} and \textit{cis}-HCOOH \mbox{($\upnu$ = 7)} in
the S$_{0}$ electronic ground-state, to the first 200 vibrational levels of the S$_{1}$ excited-state. The use
of different geometries in the two electronic states allows us to approximately reproduce the
experimental frequencies \citep{Beaty-Travis_2002}. The absorption spectrum is obtained using the transition dipole
moments obtained for \mbox{geom-S$_{0}$}.

The calculated radiative lifetimes of the different vibrational levels of the S$_{1}$ electronic excited-state vary from \mbox{75 $\times$ 10$^{-6}$ s} to \mbox{375 $\times$ 10$^{-6}$ s}, but each level has a different probability to decay towards the \textit{trans} of \textit{cis} well of the S$_{0}$ ground electronic state. We explicitly determine the probability to fluoresce into each conformer by calculating:

\begin{equation}
P_{cis\,(\nu)}=\sum_{\nu',\,cis} \left| \,<f_{\nu'}^{S_{0}}\,|\,d\,|\,f_{\nu}^{S_{1}}>\, \right|^{2}
\end{equation}

\noindent and

\begin{equation}
P_{trans\,(\nu)}=\sum_{\nu',\,trans} \left| \,<f_{\nu'}^{S_{0}}\,|\,d\,|\,f_{\nu}^{S_{1}}>\, \right|^{2}
\end{equation}

\noindent where we separate the contributions of the $\upnu'$ levels corresponding to the \textit{trans} or \textit{cis} conformers and normalize the sum to 1. We then normalize the above values and compute
P$_{cis}$($\upnu$)/(P$_{cis}$($\upnu$)+P$_{trans}$($\upnu$)) 
and 
P$_{trans}$($\upnu$)/(P$_{cis}$($\upnu$)+P$_{trans}$($\upnu$)) for $\upnu$ levels corresponding to absorption
energies below about 40000\,cm$^{-1}$ \mbox{($E$ $<$ 5 eV)}, approximately the energy for which the dominant
photodissociation channel opens and fluorescence starts to become negligible.

In summary, with these \textit{ab initio} calculations we estimate the \textit{cis}- and \textit{trans}-HCOOH
cross-sections $\upsigma_{\uplambda_{\rm i}}$ for absorption of photons with energies lower than about 40000\,cm$^{-1}$ (those producing fluorescense). These absorptions radiatively excite the molecule to the S$_{1}$ electronic
excited-state. We explicitly compute the $\upsigma_{\uplambda_{\rm i}}$ values for each photon energy as well as the
probabilites to fluoresce back to a specific  \textit{cis} or \textit{trans} state (i.e. we determine the normalized
probabilities of the different \textit{trans $\rightarrow$ cis}, \textit{trans $\rightarrow$ trans}, \textit{cis $\rightarrow$ cis}, \textit{cis $\rightarrow$ trans} bound-bound transitions). The  $\upsigma_{\uplambda_{\rm i}}$(\textit{trans}) and $\upsigma_{\uplambda_{\rm i}}$(\textit{cis}) cross-sections and the $P_{trans \rightarrow cis}$ and $P_{cis \rightarrow trans}$ probabilities are plotted in Fig.~\ref{fig:cross_extin}.

\section{Estimation of the photoisomerization rate in the Orion Bar}\label{photoisomerization_rate}  

The number of photoisomerizations per second depends on the flux of FUV photons with
energies below 5\,eV. The \textit{trans}-to-\textit{cis} and \textit{cis}-to-\textit{trans} photoisomerization rates ($\upxi_{\rm tc}$ and $\upxi_{\rm ct}$) are derived from the discrete sums:

\begin{equation}
\upxi_{\rm tc} = \sum\nolimits_{\uplambda_{\rm i}} N_{\rm ph,\,\uplambda_{\rm i}} \cdot \upsigma_{\uplambda_{\rm i}} (trans) \cdot P_{t \rightarrow c}
\end{equation}

\noindent and

\begin{equation}
\upxi_{\rm ct} = \sum\nolimits_{\uplambda_{\rm i}} N_{\rm ph,\,\uplambda_{\rm i}} \cdot \upsigma_{\uplambda_{\rm i}} (cis) \cdot P_{c \rightarrow t}
\end{equation}

\noindent where $\upsigma_{\uplambda_{\rm i}}$ is the absorption cross-section from a given conformer (in cm$^{2}$ photon$^{-1}$) and $P$ is
the probability to fluoresce from one isomer to the other. Both quantities are determined from
our \textit{ab initio} calculations (previous section). $N_{\rm ph,\,\uplambda_{\rm i}}$ (photon cm$^{-2}$ s$^{-1}$) is the flux of photons at
each wavelength producing absorption.

In order to estimate the most realistic $\upxi_{\rm tc}$ and $\upxi_{\rm ct}$ rates for the FUV-irradiation conditions in the Orion Bar, we used the Meudon PDR code \citep{LePetit_2006} and calculate
\mbox{$N_{\rm ph}(\uplambda)$} at different cloud depth $A_{\rm V}$ values. Following our previous studies of the Bar \citep{Cuadrado_2015a,Goicoechea_2016} we run a model of an isobaric PDR \mbox{($P_{\rm th}$/k = 10$^{8}$ K cm$^{-3}$)} illuminated by \mbox{$\chi$ = 4 $\times$ 10$^{4}$} times the mean
interstellar radiation field \citep{Draine_1978}. For photons in the \mbox{$\uplambda$ = 2000 $-$ 3000 \AA} range, we adopt
  \mbox{$N_{\rm ph}(\uplambda)$ = 4 $\times$ 10$^4$ $\times$ 732 $\times$ $\uplambda^{0.7}$} photon cm$^{-2}$ s$^{-1}$ $\AA^{-1}$ at the PDR edge \mbox{($A_{\rm V}$ = 0)} \citep{vanDishoeck_1982}. We use a constant
dust composition and size distribution  that reproduces standard
interstellar extinction curve \citep{Cardelli_1989}.

Table~\ref{Table_photo_rates} shows the resulting photoisomerization rates at different cloud depths, the
expected \textit{trans}-to-\textit{cis} HCOOH abundance ratio at equilibrium, and the time needed to reach the
equilibrium ratios (neglecting photodissociation).

 \begin{table*}
 \centering 
 \caption{Photoisomerization rates for the irradiation conditions in the Orion Bar.}
 \label{Table_photo_rates}   
 
\begin{tabular}{c c c c c c@{\vrule height 10pt depth 5pt width 0pt}}     
 \hline\hline

Cloud depth          &      $N_{\rm ph}$ (2300 $-$ 2800 $\AA$)  &      $\upxi_{\rm tc}$     &   $\upxi_{\rm ct}$    &    \textit{trans}/\textit{cis}-HCOOH     &     Time          \\
$A_{\rm V}$ [mag]    &      [photons cm$^{-2}$ s$^{-1}$]       &      [s$^{-1}$]           &   [s$^{-1}$]          &    ratio at equilibrium         &     [years]       \\
\hline
0               &      1.96 $\times$ 10$^{12}$             &      5.45 $\times$ 10$^{-11}$   &   1.27 $\times$ 10$^{-10}$   &    2.3                   &     1.7 $\times$ 10$^{2}$         \\
1               &      1.80 $\times$ 10$^{11}$             &      3.41 $\times$ 10$^{-12}$   &   9.89 $\times$ 10$^{-12}$   &    2.9                   &     2.4 $\times$ 10$^{3}$         \\
2               &      2.84 $\times$ 10$^{10}$             &      3.79 $\times$ 10$^{-13}$   &   1.31 $\times$ 10$^{-12}$   &    3.5                   &     1.9 $\times$ 10$^{4}$         \\
3               &      4.93 $\times$ 10$^{9}$              &      4.62 $\times$ 10$^{-14}$   &   1.87 $\times$ 10$^{-13}$   &    4.1                   &     1.3 $\times$ 10$^{5}$         \\
5               &      1.69 $\times$ 10$^{8}$              &      7.83 $\times$ 10$^{-16}$   &   4.29 $\times$ 10$^{-15}$   &    5.5                   &     6.2 $\times$ 10$^{6}$         \\
         \hline
         \end{tabular}  
           \end{table*}

The use of constant dust grain properties through the PDR is likely the most important
simplification for the calculation of the photoisomerization rates $\upxi_{\rm tc}$ and $\upxi_{\rm ct}$. Grain populations are known to evolve in molecular clouds, especially in PDRs where the sharp attenuation of a
strong FUV field results in a stratification of the dust and PAH properties with $A_{\rm V}$ \citep{Draine_2003}.
Therefore, although the varying optical properties of grains are difficult to quantify and include
in PDR models, they likely play a role on how FUV photons of different energies are
differentially absorbed as a function of $A_{\rm V}$ \citep{Goicoechea_2007}. For the particular case of HCOOH, the strength
and width of the 2175\,$\AA$ extinction bump \citep{Cardelli_1989} naturally divides the range of photons producing
HCOOH photodissociation (those with \mbox{$E$ $>$ 5 eV}) from those producing fluorescence (\mbox{$E$ $<$ 5 eV}).
The extinction bump has been related with the absorption of FUV photons by PAH mixtures and small carbonaceous grains \citep{Joblin_1992,Draine_2003}. Although it is not known how the bump evolves with $A_{\rm V}$, it clearly
determines how the lower-energy FUV photons are absorbed. In Fig.~\ref{fig:cross_extin} (bottom panel) we
show different extinction curves for different PAH abundances \citep{Goicoechea_2007}. Optical properties are taken
from \citet{Li_2001} and references therein. In addition, and as in most PDR models, our
predicted FUV spectrum does not include the absorption produced by hundreds of molecular
electronic transitions blanketing the FUV spectrum (other than H$_2$ and CO lines). All together,
our assumption that the detected \textit{cis}-HCOOH arises from PDR layers in which the flux of
photons with \mbox{$\uplambda$ $>$ 2500 $\AA$} dominates over the higher-energy photodissociating photons is very
plausible.

\section{Alternative mechanisms for \textit{trans}-to-\textit{cis} isomerization in the ISM}\label{alternative_mechan}

Searching for further support to the photoswitching scenario, we qualitatively explored
other possibilities that may apply in interstellar conditions. In the laboratory, \textit{trans}-to-\textit{cis}
isomerization has been observed in molecular ices irradiated by near-IR photons \citep{Macoas_2004,Olbert-Majkut_2008}.
Hence, isomerization of solid HCOOH and subsequent desorption to the gas-phase might also be
responsible of the \textit{cis}-HCOOH enhancement. However, owing to the short lifetime of the \textit{cis}
conformer observed in ices (a few minutes if the irradiation is stopped, \citealt{Macoas_2004}), a very strong flux of
IR photons would be needed to maintain significant abundances of solid \textit{cis}-HCOOH. In
addition, near-IR photons penetrate molecular clouds much deeper than FUV photons, and
one would have expected to detect \textit{cis}-HCOOH in all positions of the Bar, and towards the Orion
hot core, a region irradiated by intense IR fields.
Alternatively, the \textit{trans}-to-\textit{cis} isomerization might be triggered by collisions with
electrons. Electrons are relatively abundant in FUV-irradiated environments (with ionisation
fractions up to about \mbox{$n_{\rm e}$/$n_{\rm H}$ $\approx$ 10$^{-4}$)} compared to shielded cloud interiors \mbox{($n_{\rm e}$/$n_{\rm H}$ $\approx$ 10$^{-9}$)}. Simple
calculations show that electrons with energies of about 0.5\,eV would be needed to overcome the
energy barrier to HCOOH isomerization, and to produce a \textit{trans}-to-\textit{cis} abundance ratio of
about 3. Such suprathermal electrons could be provided by the photoionisation of low ionisation
potential atoms (C, S, Si...), but their abundance sharply decrease with $A_{\rm V}$ \citep{Hollenbach_1999}. We estimate that
at a cloud depth of \mbox{$A_{\rm V}$ = 2 mag}, HCOOH collisional isomerization, if effective, could
compete with photoswitching only if the elastic collisional rate coefficients were very high, of
the order of \mbox{$>$10$^{-6}$ cm$^{3}$ s$^{-1}$} for a typical electron density of \mbox{$n_{\rm e}$ $<$ 1 cm$^{-3}$} in PDRs. However, the
detection of \textit{trans}-HCOOH, but not \textit{cis}-HCOOH, towards other PDRs such as the Horsehead \citep{Guzman_2014},
with similar electron densities but much lower FUV photon flux ($>$100 times less than the Bar),
supports a photoswitching mechanism in the Orion Bar (i.e. high $\upxi_{\rm tc}$ and $\upxi_{\rm ct}$ rates), but makes it
too slow for the Horsehead and other low FUV-flux sources. Either way, we encourage
laboratory and theoretical studies of the possible role of electron collisions, as well as of a more
detailed investigation of the HCOOH, and other species, photoswitching mechanism.

\begin{figure*}[ht]
\centering
\includegraphics[scale=0.56,angle=0]{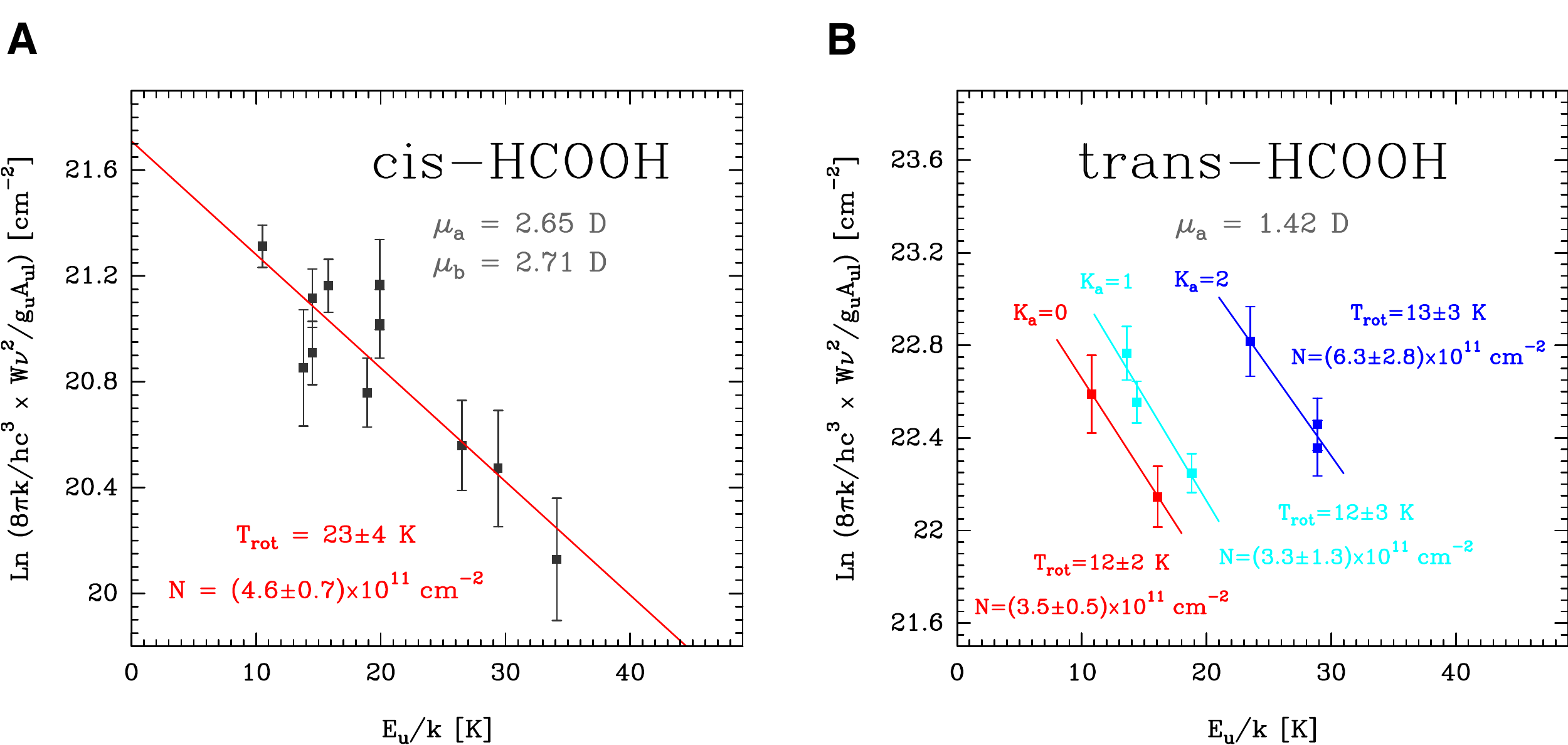}
\caption{Rotational population diagrams from the observed HCOOH lines towards
the Orion Bar, \mbox{(+10$''$, -10$''$)} position. \textit{Left}: Diagram for the \textit{cis} conformer (measurements lie along a single
component). \textit{Right}: Diagram for the \textit{trans} conformer showing how different $K_{\rm a}$ rotational
ladders split in different components. Fitted values of the rotational temperature, $T_{\rm rot}$, and column
density, $N$,  are indicated in each panel (see also Table~\ref{Table_results}).}
\label{fig:rot_diag}
\end{figure*}

\begin{table*}[!ht]
 \centering 
 \caption{Rotational temperatures ($T_{\rm rot}$), column densities ($N$), and abundances towards the Orion Bar PDR, \mbox{(+10$''$,-10$''$)} position.}
 \label{Table_results}     
  	
 \begin{tabular}{l c c c c c c c@{\vrule height 10pt depth 5pt width 0pt}}     
 \hline\hline

  & \multicolumn{2}{c}{Extended source} \rule[0.15cm]{0cm}{0.2cm}\ &  &  \multicolumn{2}{c}{Semi-extended source} \rule[0.2cm]{0cm}{0.2cm}\ & &  \\ \cline{2-3} \cline{5-6}

  & $T_{\rm rot}$ [K] & $N$(X) [cm$^{-2}$] &  & $T_{\rm rot}$ [K]  & $N$(X) [cm$^{-2}$] & Abundance$^{*}$  &  Notes \rule[0.4cm]{0cm}{0.1cm}\ \\

 \hline  
 
  $cis$-HCOOH                       &   23 (4)    &  4.6 (0.7) $\times$ 10$^{11}$  &  &     21 (4)   &      4.2 (0.6) $\times$ 10$^{12}$   &                                    &     a        \\
  \hline
  $trans$-HCOOH $K_{\rm a}$ = 0       &   12 (2)    &  3.5 (0.5) $\times$ 10$^{11}$  &  &     6 (1)    &      4.1 (0.6) $\times$ 10$^{12}$   &                                    &     a, b     \\
  $trans$-HCOOH $K_{\rm a}$ = 1       &   12 (3)    &  3.3 (1.3) $\times$ 10$^{11}$  &  &     6 (1)    &      3.6 (2.1) $\times$ 10$^{12}$   &                                    &     a        \\
  $trans$-HCOOH $K_{\rm a}$ = 2       &   13 (3)    &  6.3 (2.8) $\times$ 10$^{11}$  &  &     7 (1)    &      5.0 (2.4) $\times$ 10$^{12}$   &                                    &     a        \\
  \hline
  \textbf{[(\textit{cis+trans})-HCOOH]}     &   ---       &  1.8 (0.3) $\times$ 10$^{12}$  &  &     ---      &      1.7 (0.3) $\times$ 10$^{13}$   &    (0.3 $-$ 2.7) $\times$ 10$^{-10}$   &     c         \\
 
  \hline
                                                 
   \end{tabular}                                                     
  \tablefoot{
  $^{*}$ The abundance of each species with respect to H nuclei is given by ${\frac{N(X)}{N_H}  =  \frac{N(X)}{N(H)+2N(H_{2})} }$, 
  with \mbox{$N$(H$_{2}$) $\simeq$ 3 $\times$ 10$^{22}$ cm$^{-2}$} and 
  \mbox{$N$(H) $\simeq$ 3 $\times$ 10$^{21}$ cm$^{-2}$} \citep[][and references therein]{Cuadrado_2015a}.
  (a) Rotational temperatures and column densities from rotational diagram analysis. 
  (b) $\Delta$$N$ calculated assuming an error of 15$\%$. 
  (c) Total column densities calculated as the sum of the \textit{cis} and \textit{trans} species.}
   \end{table*}

 \section{Rotational diagrams and column density calculation} \label{DR}                 

Owing to the large number of detected HCOOH lines, we calculated rotational temperatures
($T_{\rm rot}$) and column densities ($N$) from rotational population diagrams. The standard relation for the
rotational diagram \citep{Goldsmith_1999} is:

\begin{equation}
{{\rm ln} \, \frac{N_{\rm u}}{g_{\rm u}}}={\rm ln} \, N-{\rm ln} \, Q_{_{T_{\rm rot}}}- \frac{E_{\rm u}}{\,{\rm k}  T_{\rm rot}} \, ,\label{eq:DR}
\end{equation}
	
\noindent with \mbox{$N_{\rm u}/g_{\rm u}$} given by

\begin{equation}
{\frac{N_{\rm u}}{g_{\rm u}}}={\frac{8\, {\rm \pi} \, {\rm k}}{{\rm h\,c^{3}}}} \cdot {\frac{\upnu_{\rm ul} ^{2}}{A_{\rm ul} \, g_{\rm u} }} \cdot \upeta_{_{\rm bf}}^{-1} \cdot \displaystyle{\int}T_{_{\rm MB}}dv  \, \, \, \, \, \, \, \, \, [cm^{-2}],
\end{equation}

In the above relation, $N_{\rm u}$ is the column density of the upper level in the optically thin limit [cm$^{-2}$], $N$ is the total column density [cm$^{-2}$], g$_{\rm u}$ is the statistical weight of the upper state of each level, $Q_{T_{\rm rot}}$ is the rotational partition function evaluated at a rotational temperature $T_{\rm rot}$, $A_{\rm ul}$ is the
Einstein coefficient [s$^{-1}$], $E_{\rm u}$/k is the energy of the upper level of the transition [K], $\upnu \mathrm{_{ul}}$ is the frequency of the \mbox{$u \rightarrow l$} transition [s$^{-1}$], \mbox{$\int T_{\rm MB}$dv} is the velocity-integrated line intensity corrected from beam efficiency \mbox{[K km s$^{-1}$]}, and $\upeta_{\rm bf}$ is the beam filling factor.
Assuming that the emission source has a 2D Gaussian shape, $\upeta_{\rm bf}$ is equal to \mbox{$\mathrm{\upeta_{bf}=\uptheta_{S}^{\, 2}/\,(\uptheta_{S}^{\, 2}+\uptheta_{B}^{\, 2})}$}, with $\uptheta_{\rm B}$ the HPBW of the telescope [arcsec] and $\uptheta_{\rm S}$ the diameter of the Gaussian source [arcsec]. The values for $\upnu_{\rm ul}$, \mbox{$E_{\rm u}$/k}, $g_{u}$, and $A_{\rm ul}$ are taken from the MADEX spectral catalogue.

Rotational diagrams were built considering two limiting cases: (i) that the detected
HCOOH emission is extended, with \mbox{$\upeta_{\rm bf}$ = 1}; and (ii) that the emission is semi-extended, with
\mbox{$\uptheta_{\rm S}$ = 9$''$} \citep{Cuadrado_2015a}. In a plot of ln($N_{\rm u}$/$g_{\rm u}$) \textit{versus} the energy of the upper level of each rotational
transition, \mbox{$E_{\rm u}$/k}, the population distribution roughly follows a straight line with a slope \mbox{-1/$T_{\rm rot}$}.
The total column density of the molecule, $N$, is obtained from the $y$-intercept and the partition
function. Figure~\ref{fig:rot_diag} shows the resulting rotational diagrams assuming extended emission. Table~\ref{Table_results} lists the $T_{\rm rot}$ and $N$ obtained by linear least squares fits. The uncertainties shown in Table~\ref{Table_results} indicate the uncertainty obtained in the fit. The uncertainties obtained in the determination of the
fit line parameters with CLASS are included in the error bars at each point of the rotational
diagram.

\begin{figure}[!ht]
\centering
\includegraphics[scale=0.4,angle=0]{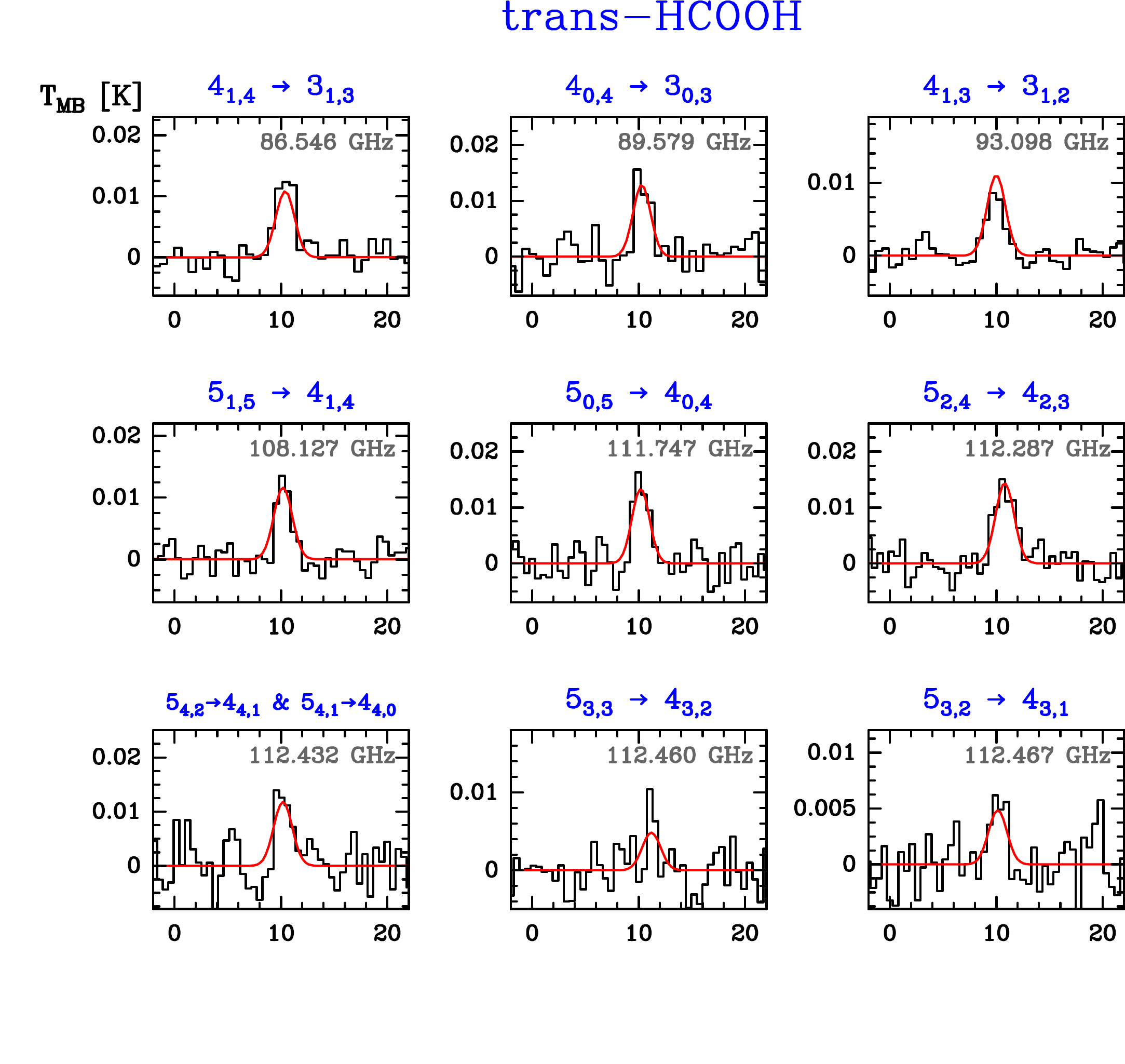}\vspace{-0.6cm}
\includegraphics[scale=0.4,angle=0]{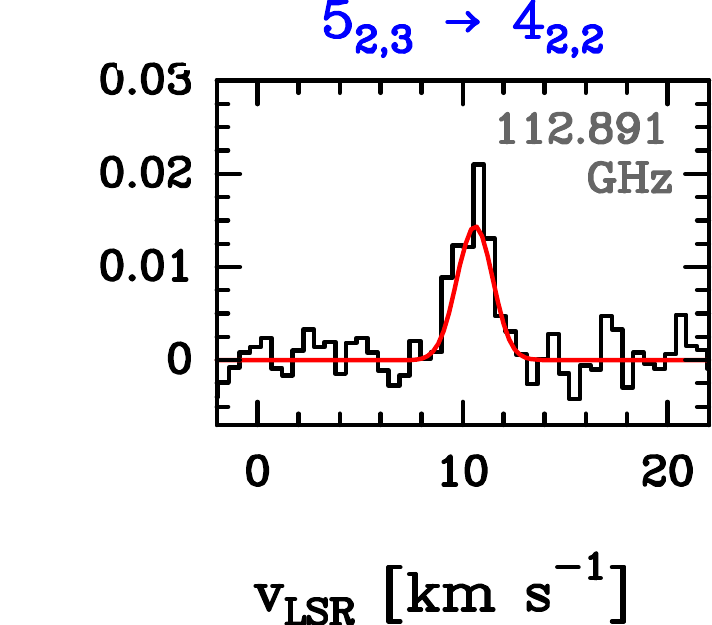}\hspace{-0.6cm}
\caption{Detected \textit{trans}-HCOOH rotational lines towards the edge of the Bar, \mbox{(+10$''$, -10$''$)} position. The ordinate
refers to the intensity scale in main beam temperature units, and the abscissa to the LSR velocity.
Line frequencies (in GHz) are indicated at the top-right of each panel together with the rotational
quantum numbers (in blue). The red curve shows an excitation model that reproduces the
observations. \textit{Cis}-HCOOH lines are shown in Fig.~\ref{fig:cis_OB}.}
\label{fig:trans_OB}
\end{figure}

To crosscheck that the relative intensities of the detected \textit{cis}- and \textit{trans}-HCOOH rotational
lines are those expected according to their inferred rotational temperatures (i.e. that the assigned
lines are not blended with lines from other molecules), we carried out a simple excitation and
radiative transfer calculation using MADEX. We assumed that the \textit{cis}- and \textit{trans}-HCOOH
rotational levels are populated following a Boltzmann distribution at a single rotational
temperature (obtained from the rotational diagrams). For a given column density $N$, the model
computes each line opacity (optically thin for the observed HCOOH lines) assuming a Gaussian
line profile (for a linewidth of \mbox{2 km s$^{-1}$}) and simulates the output mm spectrum at a given
spectral resolution. Figures~\ref{fig:cis_OB} and \ref{fig:trans_OB} show the observed spectra (black histograms) and the
modelled lines (red curves) for the $T_{\rm rot}$ and $N$ values obtained assuming extended emission. The
good agreement of the fits, and lack of any other candidate line from a different molecule in our
catalogue, confirms that all detected lines belong to \textit{cis}- and \textit{trans}-HCOOH.

\section{Non-detection of \textit{cis}-HCOOH towards the Orion BN/KL hot core and Barnard-B1} \label{non-detection}    

\begin{figure}[t]
\centering
\includegraphics[scale=0.4,angle=0]{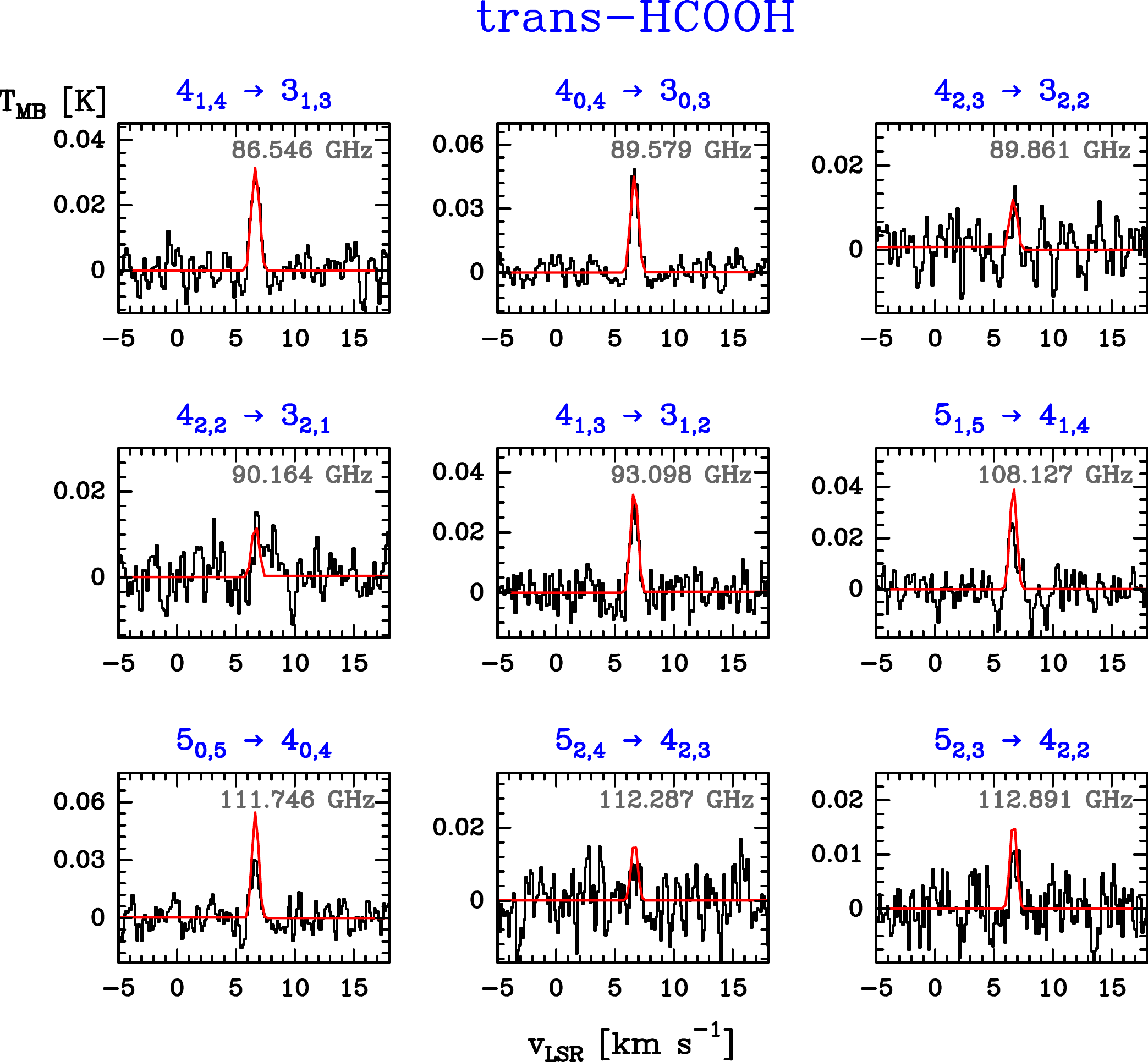}\\
\vspace{0.4cm}
\includegraphics[scale=0.4,angle=0]{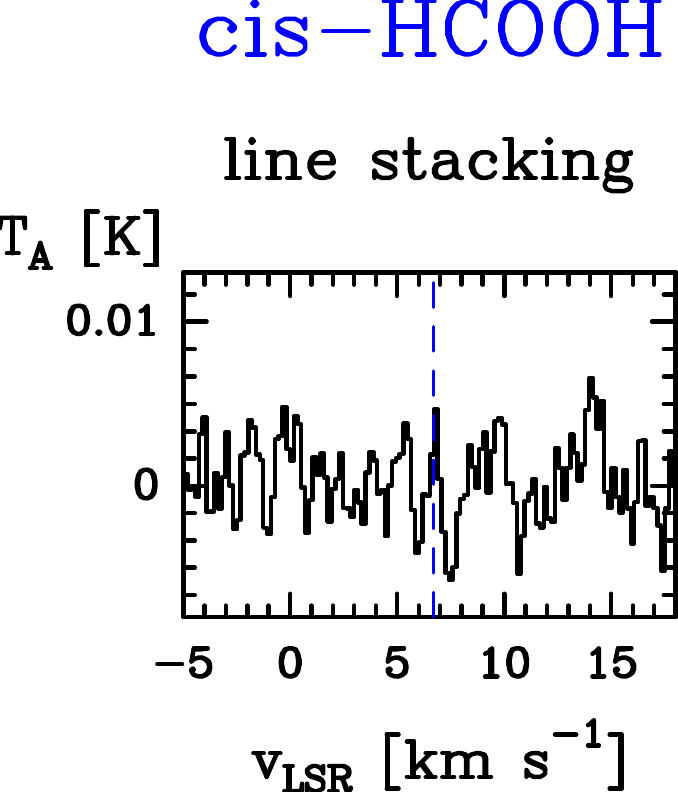}
\caption{Detected \textit{trans}-HCOOH rotational lines towards the cold cloud Barnard 1-b. The
ordinate refers to the intensity scale in main beam temperature units and the abscissa to the
Doppler velocity. Line frequencies (in GHz) are indicated at the top of each panel together with
the rotational quantum numbers. The red curve shows an excitation model that reproduces the
rotational population diagram. The bottom panel shows the stacked spectra for \textit{cis}-HCOOH.}
\label{fig:trans_cis_B1}
\end{figure}

We searched for \textit{cis}-HCOOH in regions shielded from strong FUV radiation fields. We
selected chemically rich sources for which we have also carried out deep mm-line surveys with
the IRAM-30m telescope. In particular, we searched for HCOOH towards the hot core in
Orion BN/KL \citep{Tercero_2010} and towards the quiescent dark cloud Barnard 1-b (B1-b; \citealt{Cernicharo_2012b}). Although we
clearly detected lines from \textit{trans}-HCOOH towards both sources, we did not find lines from \textit{cis}-HCOOH above the detection limit of these deep surveys. Using the MADEX excitation code, we
derived lower limits to the \textit{trans}-to-\textit{cis} abundance ratio towards these sources. Below we
summarise the main results from these observations:

{\bf Orion BN/KL hot core}: the hot core is embedded in the Becklin-Neugebauer/Kleinmann-Low massive star-forming region, at $\sim$4$'$ North-West from the Orion Bar, and $\sim$0.5$'$ North-West
from the Trapezium stars. Relatively narrow lines \mbox{($\Delta$v$_{\rm FWHM}$ $\approx$ 3 km s$^{-1}$)} 
corresponding to $a$-type
transitions of \textit{trans}-HCOOH, with upper level energies up to \mbox{$E_{\rm u}$/k $\approx$ 300 K}, 
are detected at a
LSR velocity of $\sim$8\,km\,s$^{-1}$. The observed line parameters are consistent with emission from the
hot core itself. This is dense, $n_{\rm H}$ of a few 10$^{7}$\,cm$^{-3}$, and hot gas at about 200\,K \citep{Blake_1987,Tercero_2010}, and also
from the more extended warm gas (about 60\,K) in the ambient molecular cloud, so-called the
extended ridge \citep{Blake_1987,Tercero_2010}. Using MADEX and our accumulated knowledge of the source structure
(see \citealt{Tercero_2010,Cernicharo_2016} and references therein), we determine 
\mbox{$T_{\rm rot}$(\textit{trans}) = 100 $\pm$ 35 K} and \mbox{$N$(\textit{trans}) = (1.0 $-$ 0.3) $\times$ 10$^{15}$ cm$^{-2}$}
 in the hot core, and 
\mbox{$T_{\rm rot}$(\textit{trans}) =  40 $\pm$ 15 K} and \mbox{$N$(\textit{trans}) = (1.0 $-$ 0.3) $\times$ 10$^{14}$ cm$^{-2}$}
in the extended ridge. We note that the extended ridge is the main responsible for the observed
\textit{trans}-HCOOH line emission in the 3\,mm band. Although we obtain much higher \textit{trans}-HCOOH
column densities compared to the Orion Bar, lines from \textit{cis}-HCOOH are not detected towards the
hot core. Assuming \mbox{$T_{\rm rot}$(\textit{trans}) = $T_{\rm rot}$(\textit{cis})}, we compute a lower limit to the \textit{trans}-to-\textit{cis} abundance
ratio of $>$100 in the hot core, and $>$30 in the extended ridge.\\\\
{\bf B1-b cold cloud}: Barnard 1 is a low mass star-forming region located in the Perseus cloud.
The cold core B1-b harbours two submillimetre continuum sources (B1-bN and B1-bS)
identified as first hydrostatic core candidates \citep{Gerin_2015}, and B1b-W, an infrared source detected with
Spitzer \citep{Jorgensen_2006}. Complex organic molecules such as CH$_3$OCOH, CH$_3$SH, and CH$_3$O have been
identified \citep{Marcelino_thesis_2007,Oberg_2010,Cernicharo_2012b}. We detect nine lines from \textit{trans}-HCOOH in the 3\,mm band. A rotational
diagram provides \mbox{$T_{\rm rot}$(\textit{trans}) = 12 $\pm$ 4 K} and \mbox{$N$(\textit{trans}) = (1.5 $\pm$ 0.5) $\times$ 10$^{12}$ cm$^{-2}$}. Figure~\ref{fig:trans_cis_B1} shows
the detected lines together with our best model fit (red curve). Lines are very narrow 
\mbox{($\Delta$v$_{\rm FWHM}$ $\approx$ 0.5 km s$^{-1}$)}, consistent with emission from quiescent cold gas 
(about 20\,K) shielded from FUV
radiation. Although the inferred \textit{trans}-HCOOH column density is similar to that obtained towards
the Orion Bar, we do not detect lines from \textit{cis}-HCOOH at the noise level of the B1-b data.
Assuming \mbox{$T_{\rm rot}$(\textit{trans}) = $T_{\rm rot}$(\textit{cis})}, we determine a lower limit to the \textit{trans}-to-\textit{cis} abundance ratio of $>$60. This is similar to that of the extended molecular Ridge of Orion, but significantly higher than towards the Bar.

\clearpage

\section{Detected \textit{cis}- and \textit{trans}-HCOOH lines towards the Orion Bar PDR} \label{tables}  

      \begin{table*}
        \begin{center}
        \caption{Line parameters for $cis$-HCOOH towards the Orion Bar, \mbox{(+10$''$, -10$''$)} position.}  \label{Table_cisOB}  
        \begin{tabular}{ c c r c c c c c c c c@{\vrule height 10pt depth 5pt width 0pt}}     
        \hline\hline       
        
   Transition & Frequency &  $E_{\rm u}$ &  $A_{\rm ul}$ &  $S_{\rm ij}$ &  $g_{\rm u}$ &  $\displaystyle{\int} T_{_{\rm MB}}$d$v$ &  $v_{_{\rm LSR}}$ &  $\Delta v$ &  $T_{_{\rm MB}}$ & S/N \rule[-0.3cm]{0cm}{0.8cm}\ \\ \cline{1-1}

      $(J_{K_{\rm a},K_{\rm c}})_{\rm u} \rightarrow (J_{K_{\rm a},K_{\rm c}})_{\rm l}$  & [MHz] & [K] & [$\mathrm{s^{-1}}$] & & & [$\mathrm{mK\, km\, s^{-1}}$] & [$\mathrm{km\, s^{-1}}$] & [$\mathrm{km\, s^{-1}}$]  & [$\mathrm{mK}$]  & \\
       
        \hline  
        
4$_{1,3}\rightarrow$ 4$_{0,4}$        &      82740.491   &    14.5  &   2.32 $\times$ 10$^{-5}$  &  4.3  &    9    &    18.9(2.3)     &     10.5(0.2)  &  1.8(0.4)   &   8.7   &   4.7   \\
4$_{1,4}\rightarrow$ 3$_{1,3}$        &      85042.744   &    13.8  &   2.10 $\times$ 10$^{-5}$  &  3.7  &    9    &    15.3(3.5)     &     10.6(0.3)  &  2.1(0.5)   &   6.6   &   3.5   \\
5$_{1,4}\rightarrow$ 5$_{0,5}$        &      86556.490   &    19.9  &   2.59 $\times$ 10$^{-5}$  &  5.1  &   11    &    26.3(3.5)     &     10.0(0.2)  &  2.4(0.4)   &   10.1  &   5.6   \\
4$_{0,4}\rightarrow$ 3$_{0,3}$        &      87694.689   &    10.5  &   2.45 $\times$ 10$^{-5}$  &  4.0  &    9    &    26.6(2.3)     &     10.6(0.1)  &  2.0(0.2)   &   12.7  &   9.6   \\
4$_{1,3}\rightarrow$ 3$_{1,2}$        &      90661.090   &    14.5  &   2.54 $\times$ 10$^{-5}$  &  3.7  &    9    &    21.2(2.4)     &     10.3(0.1)  &  1.8(0.3)   &   11.2  &   4.8   \\
7$_{0,7}\rightarrow$ 6$_{1,6}$        &      90910.082   &    29.4  &   1.50 $\times$ 10$^{-5}$  &  3.5  &   15    &    10.9(2.4)     &     10.1(0.2)  &  1.5(0.4)   &   6.9   &   3.9   \\
6$_{1,5}\rightarrow$ 6$_{0,6}$        &      91291.549   &    26.5  &   2.95 $\times$ 10$^{-5}$  &  5.9  &   13    &    20.1(3.5)     &     10.3(0.2)  &  2.5(0.6)   &   7.5   &   5.7   \\
7$_{1,6}\rightarrow$ 7$_{0,7}$        &      97025.449   &    34.1  &   3.42 $\times$ 10$^{-5}$  &  6.6  &   15    &    15.5(2.4)     &     10.5(0.1)  &  1.8(0.3)   &   7.0   &   4.4   \\
8$_{1,7}\rightarrow$ 8$_{0,8}$        &     103845.157   &    42.8  &   4.02 $\times$ 10$^{-5}$  &  7.1  &   17    &    20.5(3.6)     &     10.0(0.2)  &  1.6(0.2)   &   12.2  &   5.2   \\
5$_{1,5}\rightarrow$ 4$_{1,4}$        &     106266.589   &    18.9  &   4.28 $\times$ 10$^{-5}$  &  4.8  &   11    &    22.2(2.4)     &     10.7(0.1)  &  1.8(0.3)   &   11.0  &   4.5   \\
5$_{0,5}\rightarrow$ 4$_{0,4}$        &     109470.705   &    15.8  &   4.87 $\times$ 10$^{-5}$  &  5.0  &   11    &    35.7(3.6)     &     10.3(0.1)  &  1.8(0.2)   &   18.6  &   7.1   \\
5$_{1,4}\rightarrow$ 4$_{1,3}$        &     113286.704   &    19.9  &   5.19 $\times$ 10$^{-5}$  &  4.8  &   11    &    39.6(5.3)     &     10.7(0.2)  &  1.9(0.3)   &   19.7  &   6.0   \\

       \hline                                                                                
       \end{tabular}  
       \end{center} 
       \end{table*}

      \begin{table*}
        \begin{center}
        \caption{Line parameters for $trans$-HCOOH towards the Orion Bar, \mbox{(+10$''$, -10$''$)} position.}  \label{Table_transOB}  
        \begin{tabular}{ c c r c c c c c c c c@{\vrule height 10pt depth 5pt width 0pt}}     
        \hline\hline       
        
   Transition & Frequency &  $E_{\rm u}$ &  $A_{\rm ul}$ &  $S_{\rm ij}$ &  $g_{\rm u}$ &  $\displaystyle{\int} T_{_{\rm MB}}$d$v$ &  $v_{_{\rm LSR}}$ &  $\Delta v$ &  $T_{_{\rm MB}}$ & S/N \rule[-0.3cm]{0cm}{0.8cm}\ \\ \cline{1-1}

      $(J_{K_{\rm a},K_{\rm c}})_{\rm u} \rightarrow (J_{K_{\rm a},K_{\rm c}})_{\rm l}$  & [MHz] & [K] & [$\mathrm{s^{-1}}$] & & & [$\mathrm{mK\, km\, s^{-1}}$] & [$\mathrm{km\, s^{-1}}$] & [$\mathrm{km\, s^{-1}}$]  & [$\mathrm{mK}$]  & \\
       
        \hline  

   4$_{1,4}\rightarrow$ 3$_{1,3}$      &       86546.180        &      13.6   &      6.35 $\times$ 10$^{-6}$    &    3.7   &    9   &    23.8(3.5)    &     10.4(0.1)   &    1.8(0.3)    &       13.7     &     7.0      \\
   4$_{0,4}\rightarrow$ 3$_{0,3}$      &       89579.168        &      10.8   &      7.51 $\times$ 10$^{-6}$    &    4.0   &    9   &    27.9(4.7)    &     10.3(0.1)   &    1.8(0.2)    &       14.9     &     5.3      \\
   4$_{2,3}\rightarrow$ 3$_{2,2}$      &       89861.473        &      23.5   &      5.69 $\times$ 10$^{-6}$    &    3.0   &    9   &    26.4(7.0)    &     10.0(0.4)   &    2.2(0.7)    &       11.3     &     3.0      \\
   4$_{1,3}\rightarrow$ 3$_{1,2}$      &       93098.350        &      14.4   &      7.91 $\times$ 10$^{-6}$    &    3.7   &    9   &    26.3(2.4)    &     10.0(0.1)   &    2.2(0.2)    &        9.9     &     6.2      \\
   5$_{1,5}\rightarrow$ 4$_{1,4}$      &       108126.709       &      18.8   &      1.30 $\times$ 10$^{-5}$    &    4.8   &    11  &    28.7(2.4)    &     10.2(0.1)   &    1.5(0.2)    &       14.0     &     8.3      \\
   5$_{0,5}\rightarrow$ 4$_{0,4}$      &       111746.771       &      16.1   &      1.49 $\times$ 10$^{-5}$    &    5.0   &    11  &    27.9(3.6)    &     10.2(0.1)   &    1.7(0.3)    &       16.2     &     5.2      \\
   5$_{2,4}\rightarrow$ 4$_{2,3}$      &       112287.131       &      28.9   &      1.27 $\times$ 10$^{-5}$    &    4.2   &    11  &    32.2(3.6)    &     10.7(0.1)   &    2.1(0.2)    &       14.2     &     6.1      \\
   5$_{3,3}\rightarrow$ 4$_{3,2}$      &       112459.608       &      44.8   &      9.73 $\times$ 10$^{-6}$    &    3.2   &    11  &    9.1(3.6)     &     11.2(0.1)   &    0.5(0.4)    &       16.5     &     5.2      \\
   5$_{3,2}\rightarrow$ 4$_{3,1}$      &       112466.993       &      44.8   &      9.73 $\times$ 10$^{-6}$    &    3.2   &    11  &    14.0(3.6)    &     10.1(0.3)   &    1.8(0.5)    &        7.3     &     2.8      \\
   5$_{4,2}\rightarrow$ 4$_{4,1}$      &       112432.278       &      67.1   &      5.47 $\times$ 10$^{-6}$    &    1.8   &    11  &    \multirow{2}{*}[0cm]{$\Big \rangle$ 24.9(6.1)}    &     \multirow{2}{*}[-0.01cm]{10.2(0.2)}   &    \multirow{2}{*}[-0.01cm]{1.6(0.4)}    &       \multirow{2}{*}[-0.01cm]{14.4}     &     \multirow{2}{*}[-0.01cm]{3.1}      \\
   5$_{4,1}\rightarrow$ 4$_{4,0}$      &       112432.305       &      67.1   &      5.47 $\times$ 10$^{-6}$    &    1.8   &    11  &                 &                 &                &                &              \\
   5$_{2,3}\rightarrow$ 4$_{2,2}$      &       112891.429       &      28.9   &      1.29 $\times$ 10$^{-5}$    &    4.2   &    11  &    29.2(3.6)    &     10.6(0.1)   &    2.0(0.2)    &       17.6     &     7.3      \\
        
          \hline                                                  
          \end{tabular}   
          \end{center} 
          \end{table*}


\end{document}